\documentclass[twocolumn]{aastex}

\usepackage{amsmath}

\usepackage{xcolor}

\fullcollaborationName{The Friends of AASTeX Collaboration}

\begin{document}

\title{Jovian electrons in the inner heliosphere: Opportunities for Multi-spacecraft Observations and Modeling}



\author{R.~D. Strauss\altaffilmark{1,2}, N. Dresing\altaffilmark{3}, N.~E. Engelbrecht\altaffilmark{1,2}, J.~G. Mitchell\altaffilmark{4}, P. K\"uhl\altaffilmark{5}, S. Jensen\altaffilmark{5}, S. Fleth\altaffilmark{5}, B. S\'anchez-Cano\altaffilmark{6}, A. Posner\altaffilmark{7}, J.S Rankin\altaffilmark{8}, C.~O. Lee\altaffilmark{9}, J.P. van den Berg\altaffilmark{1}, S.E.S. Ferreira\altaffilmark{1}, B. Heber\altaffilmark{5}}


\altaffiltext{1}{Center for Space Research, North-West University, Potchefstroom, South Africa} 
\altaffiltext{2}{National Institute for Theoretical and Computational Sciences (NITheCS), South Africa}
\altaffiltext{3}{Department of Physics and Astronomy, University of Turku, Turku, Finland}
\altaffiltext{4}{NASA Goddard Space Flight Center, Greenbelt, USA}
\altaffiltext{5}{Institut f\"ur Experimentelle und Angewandte Physik, Christian Albrechts-Universit\"at zu Kiel, Germany}
\altaffiltext{6}{School of Physics and Astronomy, University of Leicester, Leicester, UK}
\altaffiltext{7}{NASA/HQ, Washington, DC, USA}
\altaffiltext{8}{Department of Astrophysical Sciences, Princeton University, Princeton, NJ 08540, USA}
\altaffiltext{9}{University of California at Berkeley, Berkeley, CA, USA}

\begin{abstract}

In this paper we explore the idea of using multi-spacecraft observations of Jovian electrons to measure the 3D distribution of these particles in the inner heliosphere. We present simulations of Jovian electron intensities along selected spacecraft trajectories for 2021 and compare these, admittedly qualitatively, to these measurements. Using the data-model comparison we emphasize how such a study can be used to constrain the transport parameters in the inner heliosphere, and how this can lead to additional insight into energetic particle transport. Model results are also shown along the expected trajectories of selected spacecraft, including the off-ecliptic phase of the Solar Orbiter mission from 2025 onward. Lastly, we revisit the use of historical data and discuss upcoming missions that may contribute to Jovian electron measurements. 

\end{abstract}


\section{Introduction}

\begin{figure*}
    \centering
\includegraphics[width=0.49\textwidth]{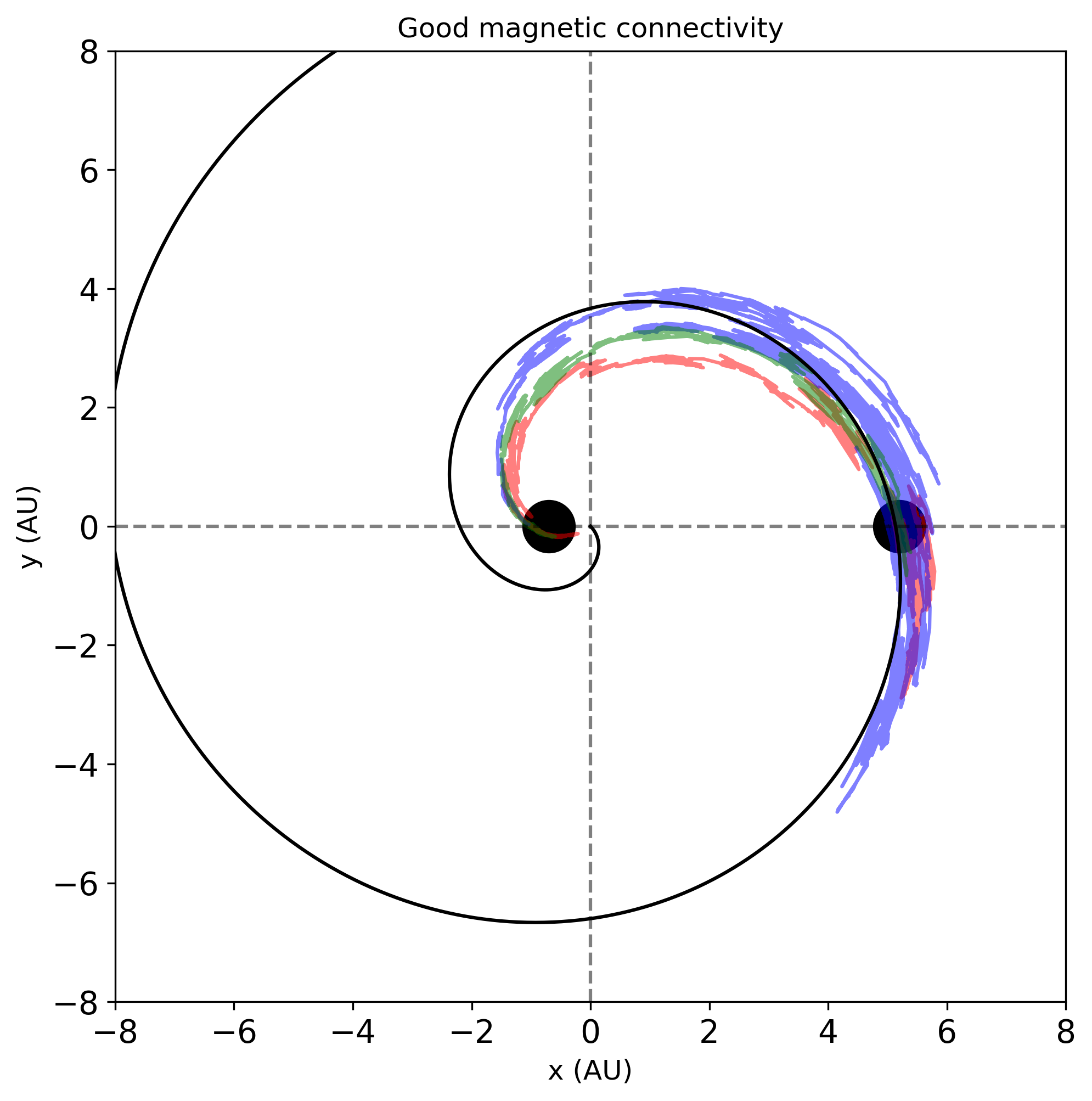}
\includegraphics[width=0.49\textwidth]{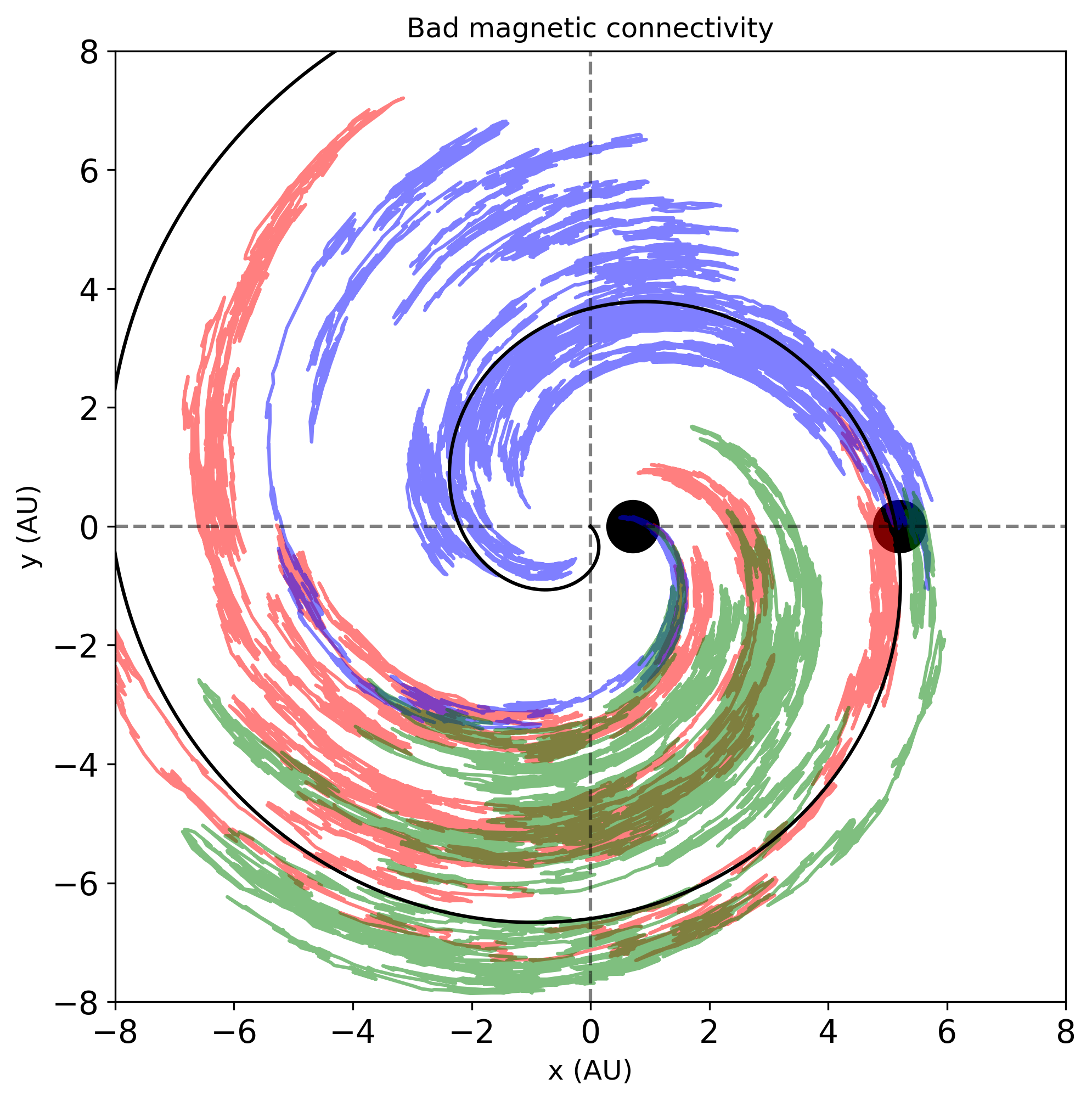}
    \caption{Examples of modelled pseudo-particle trajectories (three different realizations of the numerical scheme are indicated by different colours) of 4 MeV Jovian electrons propagating between Jupiter and Earth. The Sun is located at the origin, Jupiter at a distance of 5.2 au away from the Sun to the right (the position indicated by a filled circle), and Earth at a radial distance of 1 au from the Sun (the position again indicated by a filled circle). Simulations are performed for times of relatively good (left panel) and relative bad magnetic connection (right panel) between the planets where the spiral line indicated the Parker HMF line connected to Jupiter. The results are taken from \citet{Straussetal2011}.}
    \label{fig:particle_trajectories}
\end{figure*}

Jupiter is the main source of a few MeV quiet-time electrons in the inner heliosphere. At Earth, the intensities of these Jovian electrons are characterized by a $\sim 13$ month periodicity due to the synodic period of Jupiter \citep[e.g.][]{Chenette1980,Moses1987}. Direct measurements by the Pioneer 10 spacecraft confirmed that these particles are indeed of Jovian origin \citep[e.g.][]{Eraker1982}. This $\sim 13$ month periodicity is due to changing levels of magnetic connection between the observer (in this case near-Earth spacecraft) and Jupiter, as illustrated in Fig. \ref{fig:particle_trajectories}. When there is a good magnetic connection, particles can travel predominantly along the \citet{Parker1958} spiral. During times of bad magnetic connection (the right panel of the figure), particles need to be scattered perpendicular to the mean field to reach the observer. This is usually a much slower process than field-aligned transport (even with the addition of pitch-angle diffusion) so that particles spend much more time being adiabatically cooled during their propagation, and subsequently reach the observer with much lower intensities \citep[see eg.][and references therein]{Vogtetal2022}. \citet{Vogtetal2020} also show that the residence time (mean propagation time to Earth) of MeV Jovian electrons is about 5 and 10~days for good and poor connection, respectively. \\

The {\it Ulysses} spacecraft, with it's highly inclined trajectory \cite[e.g.][]{Heberetal1997}, measured Jovian electrons propagating to high heliospheric latitudes. Onboard {\it Ulysses} were a number of scientific instruments of which the Kiel Electron Telescope (KET) covers a wide range of electron fluxes from about 2.5 MeV to 6000 MeV \citep{Simpsonetal1992,Heberetal2001,Heberetal2002}. Assuming a quasi-constant Jovian source \citep[e.g.][]{Simpsonetal1974,Eraker1982} located in the inner heliosphere, Jovian electrons may serve as a very good probe of interplanetary transport conditions and the underlying processes influencing their transport \citep[e.g.][]{Ferreiraetal2001b}. Subsequent modelling of these particles by e.g. \cite{FerreiraEA01} and \cite{Moeketsietal2005} showed that these particles reach high latitudes due to efficient perpendicular scattering in the meridional plane. Such studies, where Jovian electron measurements are compared to simulation results, have led to a number of similar insights regarding particle transport processes through turbulent plasmas \cite[see also][]{Kissmanetal2003,Ferreiraetal2004}. Furthermore, Jovian electron observations, due to the larger parallel mean free paths and relatively drift-free transport of these particles, have been employed to investigate the structure of the heliospheric magnetic field. \citet{SternalEA11}, for instance, combined electron transport modelling with careful analyses of observations to find evidence for a \citet{Fisk96}-type heliospheric magnetic field in {\it Ulysses} count rates of 2.5--7 MeV electrons.\\

The decentral location of the Jovian source, in combination with the Parker heliospheric magnetic field geometry and the essentially drift-free propagation of these particles, provides an almost ideal opportunity for the investigation of low-energy electron diffusion coefficients in particular, through careful modelling of observed Jovian intensities. Essentially, perpendicular diffusion would be the dominant mechanism in their transport during periods of poor magnetic connectivity of the observer with the Jovian source, while parallel diffusion dominates during magnetically well-connected intervals. To date, relatively few studies have taken advantage of this phenomenon, as the majority of studies reporting on particle mean free paths focus on modelling the transport of solar energetic particles (SEPs), or galactic cosmic rays \cite[for a brief review of these, see][]{EngelbrechtEA22b}. \citet{Chenette77} do take advantage of this to provide one of the earliest observational estimates for the electron perpendicular mean free path, with subsequent studies providing refined estimates including values for the electron parallel mean free path \citep{Vogtetal2020,Vogtetal2022}, and estimates for the potential rigidity dependence of these particle's perpendicular mean free paths \citep{ZhangEA07}. \citet{Engelbrecht_2022} report on radial and rigidity dependences, as well as values at $1$~au, for electron parallel and perpendicular mean free paths which yield model results of spacecraft observations of both good and poor magnetic connectivity with Earth. These could then be compared with predictions for these quantities calculated from various scattering theories. However, it should be noted that those authors report a degeneracy in the parameter set required to fit single spacecraft observations, which may be disentangled via comparisons with data sets originating from multiple spacecraft, as described here.\\

Recently, \citet{mitchell2022} reported measurements inside of $\sim 0.5$ au made by the Parker Solar Probe (PSP) spacecraft. Although Jovian electron measurements have been reported at such small radial distances before (e.g. \citet{ErakerSimpson1979} reports Mariner 10 observations just inside 0.5 au during the Mercury encounter), the clear signal from PSP and the peculiar time profile have led to renewed interest in the propagation of Jovian electrons to the very inner heliosphere. Similar to the earlier {\it Ulysses} results, Jovian electrons from such a unique observational position, compared to extensive modelling results, can again lead to insight and potential breakthroughs regarding particle transport. Moreover, there is currently a fleet of spacecraft in the inner heliosphere that have the ability to measure Jovian electron intensities, including, PSP, STEREO A (STA), SOHO, BepiColombo (hereinafter Bepi), Solar Orbiter, and MAVEN. Such multi-spacecraft Jovian electron measurements have the {potential to provide precisely the multiple sets of observations required to provide new insights into MeV electron transport.}

In this paper we simulate the intensity of Jovian electrons along a number of these spacecraft trajectories for 2021. This time interval was still in the minimum of solar activity with only a couple of SEP events observed, mainly towards the second part of the year. These simulations are qualitatively compared to various {\it in-situ} measurements and we evaluate the ability of each spacecraft to accurately measure Jovian electrons, and also the ability of the model to reproduce the correct temporal variations. Thereafter, we perform long-term simulations of the Jovian electron intensity along a few selected spacecraft trajectories to show what type of time profiles can be expected in the future. We do the same for historical spacecraft ({\it Helios} 1 and 2, and near-Earth spacecraft) as motivation to re-analyze these older data. We close the paper with a section devoted to a discussion of upcoming missions that can contribute future Jovian electron measurements, and various avenues for future work.\\

\section{Spacecraft measurements}

In this section we briefly discuss the different spacecraft missions and instruments used to compile Jovian electron intensities for 2021. We have not filtered out contributions from SEP events, which leads to significant contamination in the second half of 2021. Additionally, the different instruments provide electron intensities at different energies, making a detailed quantitative study impractical at the moment. However, for the purposes of this initial qualitative study, these issues will only effect the results in a quantitative fashion and will therefore not affect the main conclusions of the paper.\\

\subsection{Parker Solar Probe}

As mentioned above, \cite{mitchell2022} recently reported observations of Jovian electrons observed by the High-Energy Telescope (HET) instrument in EPI-Hi within the Integrated Science Investigation of the Sun (IS$\odot$IS) instrument suite on board PSP. The HET instrument utilizes a standard dE/dx versus residual energy measurement technique to measure electrons in the energy range $\sim$0.5 - 6 MeV.  Further details on the IS$\odot$IS/EPI-Hi instruments are provided by \citet{mccomas2016} and \citet{wiedenbeck2017}. Due to the several-month orbital period, as well as the high angular (and radial) velocity and eccentricity of the PSP orbit, the spacecraft is nominally magnetically connected to Jupiter for a brief ($\sim$10-15 day) time period during each orbit. For roughly the first half of the PSP mission, these brief periods of nominal connectivity occurred as the spacecraft was leaving its closest approaches with the Sun as determined by the positions of the spacecraft and Jupiter, as well as the average solar wind velocity during that time.  \\ 

\subsection{Solar Orbiter}

The High Energy Telescope (HET) of Solar Orbiter is part of the Energetic Particle Detector \citep[EPD;][]{pacheco20} suite. It consists of two near-identical units located on the spacecraft with orthogonal viewing directions. Each HET unit is a double-sided telescope stack consisting of a central bismuth germanium oxide (BGO) scintillator (C) and two solid state detectors (SSDs) on each side. Therefore, HET is capable of covering four viewing directions in total. The four SSDs are segmented, with the outer segments acting as anticoincidence. HET utilizes the dE/dx-E method and the dE/dx-dE/dx method for stopping and penetrating particles, respectively. The HET instrument is designed to cover the energy range from a few MeV/nuc up to a few hundred MeV/nuc for ions \citep[see Fig.~3 in][]{pacheco20} and relativistic electrons to above 10 MeV. Additional information and updates are also given in \citet{wimmer-etal-2021}. 

\subsection{STEREO A}

The High Energy Telescope \citep[HET; ][]{vonRosenvinge2008} carried by both spacecraft of the STEREO mission \citep[][]{Kaiser2008} measures protons in the energy range of 13 to 100~MeV, and electrons from 0.7 to 4~MeV. We use STEREO~A/HET measurements of the highest electron energy channel E3, providing 2.8--4.0~MeV electron intensities. The instrument provides only one viewing direction, which was aligned with the nominal direction of the Parker spiral magnetic field at the start of the STEREO mission. However, since the spacecraft was put upside down after its solar superior conjunction in 2014--2015, HET points perpendicular to the nominal Parker spiral. \\
\subsection{BepiColombo}

The BepiColombo Environment Radiation Monitor (BERM) on board the BepiColombo mission \citep{Benkhoff2010} is an instrument that is part of the housekeeping suite of the European Mercury Planetary Orbiter (MPO). BERM is designed to measure the radiation environment encountered by BepiColombo during its cruise phase and after orbit insertion at Mercury, being one of the few instruments that is operated continuously since its launch in 2018 \citep{Pinto2022}. BERM measures electrons with energies from ${\sim}$170 keV to $\sim$10 MeV, protons from $\sim$1.35 MeV to $\sim$160 MeV, and heavy ions with Linear Energy Transfer from 1 to 50 MeV·mg$^{-1}$·cm$^{2}$. In this study, we use the electron channel E4 that is a integral channel sensitive to electrons with energies 1.1-$\infty$ MeV \citep{Pinto2022}.\\ 

\subsection{SOHO}

{The Electron Proton Helium INstrument (EPHIN) is part of the Comprehensive Suprathermal and Energetic Particle Analyzer \citep[COSTEP;][]{Mueller-Mellin-etal-1995} experiment aboard the SOHO spacecraft. The SOHO mission was launched in December 1995 and is in orbit around the Lagrangian point L1. EPHIN consists of 6 solid-state silicon detectors, labeled A-F from top to bottom, surrounded by a scintillator tube that serves as anti-coincidence. The energy range for stopping electrons extends from 0.2~MeV to about 10~MeV.
The data is based on the E1300 coincidence channel. Since 2017, the instrument has been operating in failure modes D and E which removes the two lowest detectors from any coincidence logic. This results in a wide energy range for the E1300. We performed a \emph{bow-tie} analysis on this channel for a more sophisticated calculation of the response factor. Assuming a power-law spectrum with spectral indices between -2.5 and -1 the effective energy of this channel is $E_{eff} = 1.8$~MeV. Also the data was filtered for SEP events as discussed in Sec.~\ref{sec:AppEphinFilterMethod} of the appendix.}

\subsection{MAVEN}

The Solar Energetic Particle (SEP) instrument on aboard the Mars Atmosphere and Volatile EvolutioN (MAVEN), which provides the energy-flux spectra of high energy particles (ions and electrons), is used \citep{Larson2015}. In particular, data from the SEP1 sensor's forward looking direction, which is nominally pointing at the Parker spiral. Here we use the integrated SEP1 electron flux in the energy range 100 -- 200 keV.\\

\section{Jovian electron transport modelling}

For the work presented here, we use the model of \citet{Straussetal2011}, rewritten by \citet{Dunzlaffetal2015}, to solve the \citet{Parker1965} transport equation and simulate the intensities of Jovian electrons in the inner heliosphere. This model was also used previously to calculate Jovian electrons propagation times \citep[][]{Straussetal2013}, a calculation later refined by \citet{Vogtetal2020}. For this model we adopt a \citet{Parker1958} heliospheric magnetic field with a nominal solar wind speed of 400 km/s. The Jovian electron source is introduced in the model by specifying the Jovian source function of \citet{vogtetal2018} at the position of Jupiter.\\

Here we use a relatively \textit{ad hoc} expressions, consistent with those used by e.g. \citet{Vogtetal2022}, to model the transport coefficients of low-energy electrons, given by 

\begin{equation}
    \lambda_{||}(r) = \frac{\lambda_0}{2} \left( 1 + \frac{r}{r_0} \right),
\end{equation}

where $\lambda_0=0.1$ au is a reference value at $r_0=1$ au. For the perpendicular mean free path it is assumed that $\lambda_{\perp} = \chi \lambda_{||}$ with $\chi = 0.015$, in an approach similar to that of \citet{FerreiraEA01}. As there is some uncertainty in these values, we also include a $25\%$ error bar, so that $\lambda_0=0.1 \pm 0.025$ au and $\chi = 0.015 \pm 0.00375$. Note that the diffusion coefficient $\kappa$ is related to the above mean free paths by $\kappa=v\lambda/3$ in the standard way \cite[see, e.g.,][]{Shalchibook}. Although these expressions are considerably simpler than those yielded by various scattering theories \cite[see, e.g.,][and references therein]{EngelbrechtEA22b}, there are several advantages implicit to their use. Primarily, they yield values at $1$~au, as well as energy and radial dependencies, consistent with what is expected from prior modelling endeavours and theory, such that the results for free parameters can, in principle, be compared to theory to draw conclusions as to the behaviour required of these coefficients so as to fit various sets of observations \cite[e.g.][]{Palmer82,BieberEA94,Droge00,Engelbrecht_2022}. A secondary advantage lies in the tractability of these expressions relative to those derived from first principles. It should be noted that, as a first approach, no solar cycle dependence is assumed in the transport coefficients employed here. Such a dependence has been inferred in the past \citep{Kanekaletal2003}, and is expected from theory, given the dependence of particle mean free paths on observed solar cycle-dependent turbulence quantities \cite[e.g.][]{ZhaoEA18,BurgerEA22}, which in turn has been demonstrated to influence the modulation of galactic cosmic ray protons \cite[e.g.][]{MolotoEngelbrecht20}. Modelling these effects, however, would require modelling electron mean free paths and the turbulence quantities they depend on from first principles \cite[e.g.][]{EngelbrechtStrauss18,Engelbrecht19}, which remains beyond the scope of the present study. {However, as shown by \cite{Kanekaletal2003}, any such time-dependent modulation effects are generally small for Jovian electrons and the changing magnetic connectivity between the Jovian source and the spacecraft remains the dominant effect. This is supported by the results of prior simulations investigating the residence times of Jovian electrons versus those of higher energy galactic cosmic ray electrons: \citet{Straussetal2011} report galactic cosmic ray electron residence times of several hundred days, as opposed to residence times of only a few days for Jovian electrons \cite[see also][]{Vogtetal2020,Vogtetal2022}. As the Jovian residence times are very small relative to the typical solar cycle-related variation times of heliospheric plasma parameters, it would be reasonable to neglect temporal variations on these time scales, and the more significant influence of the relatively shorter 13-month periodicity associated with varying magnetic connectivity could also be expected to be more significant. It should also be noted, however, that caution needs be taken when drawing conclusions as to low-energy ($\sim $ MeV) electron modulation by comparing modulation effects observed for high energy galactic cosmic ray electrons and Jovian electrons. This is due to the fact that these particles are influenced by different transport mechanisms. For instance, the transport of high energy galactic cosmic ray electrons is significantly influenced by drift effects, while that of Jovians is not \cite[see, e.g.,][]{FerreiraEA01,Engelbrecht19}. Furthermore, the diffusion coefficients of these particles also differ significantly \cite[e.g.][and references therein]{EngelbrechtEA22b}. These latter points, when considered in combination with the fact that Jovian electrons are exposed to a modulation volume considerably smaller than that experienced by galactic electrons, further support the observational evidence presented by \cite{Kanekaletal2003} for the predominant influence of magnetic connectivity on time scales short relative to solar cycle time scales on the transport of Jovian electrons.}\\

\subsection{Initial modelling results}

\begin{figure*}
    \centering
\includegraphics[width=0.99\textwidth]{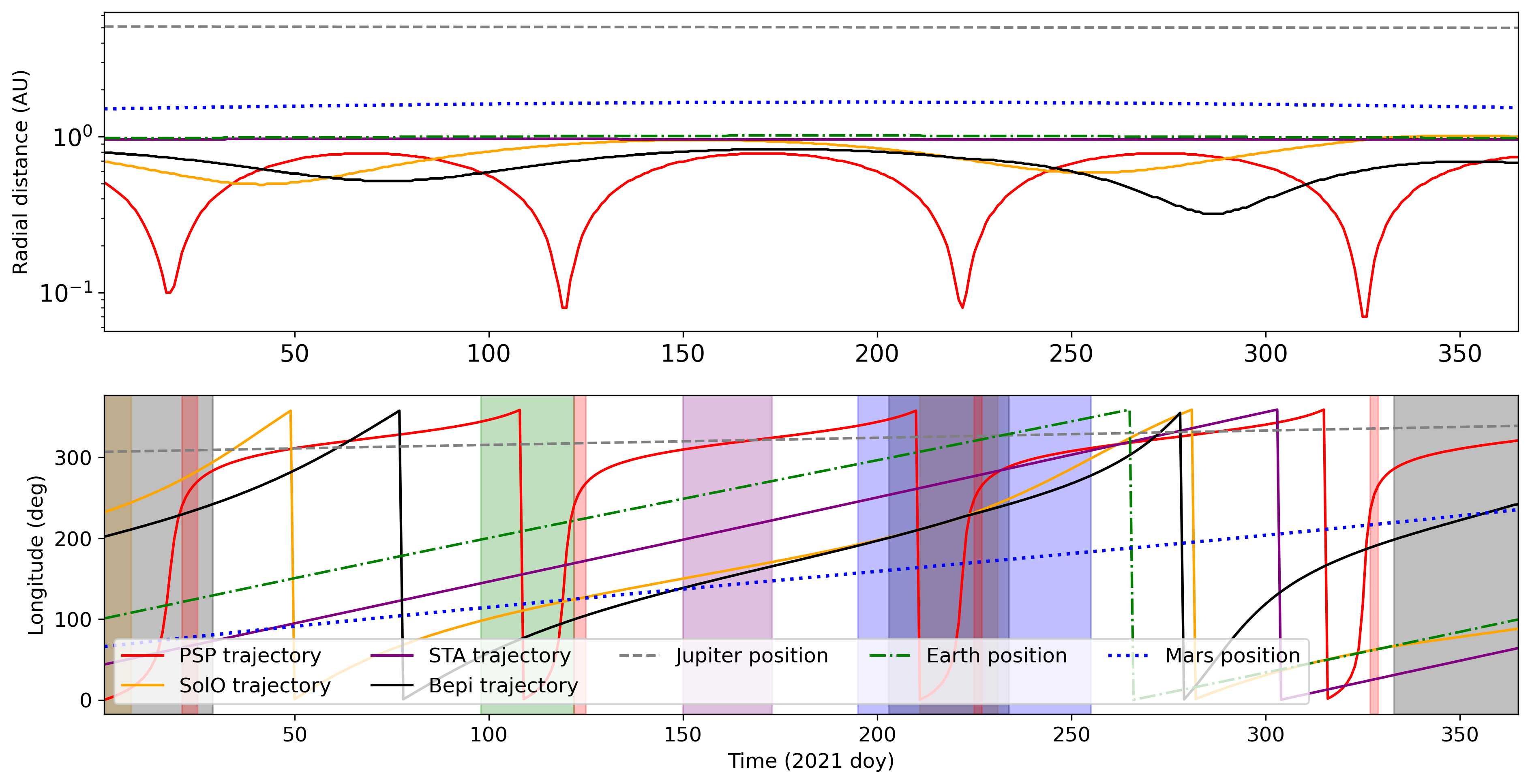}\\
\includegraphics[width=0.52\textwidth]{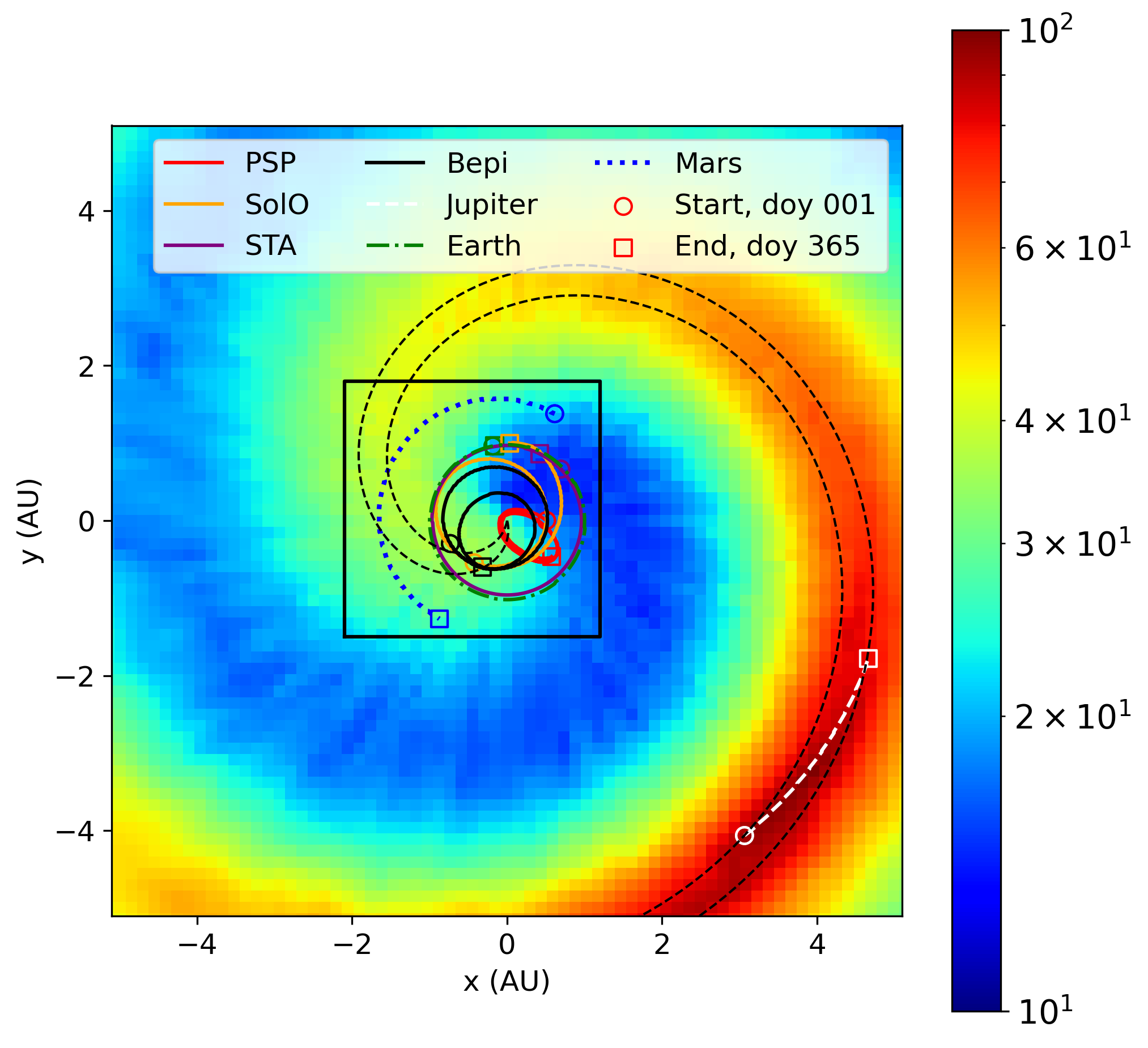}
\includegraphics[width=0.46\textwidth]{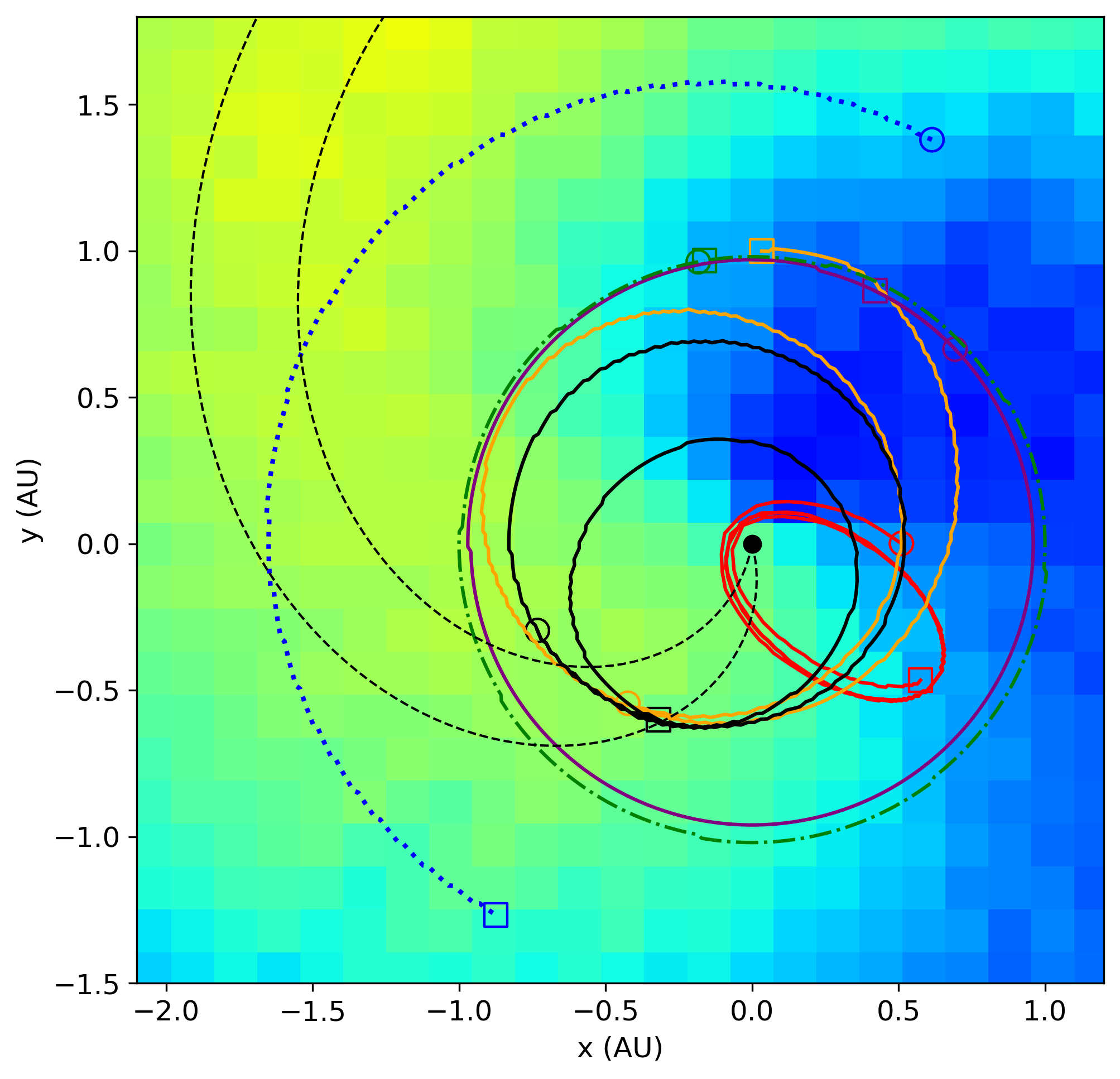}
    \caption{Selected spacecraft trajectories for 2021. In the bottom panels, these trajectories are overlaid onto simulation results for 4 MeV Jovian electrons with the Parker spiral field connected to Jupiter also shown. The colorbar shows the intensity, normalized to 100\%, of Jovian electrons on a logarithmic scale.}
    \label{fig:2D_trajectories}
\end{figure*}

\begin{figure*}
    \centering
\includegraphics[width=0.99\textwidth]{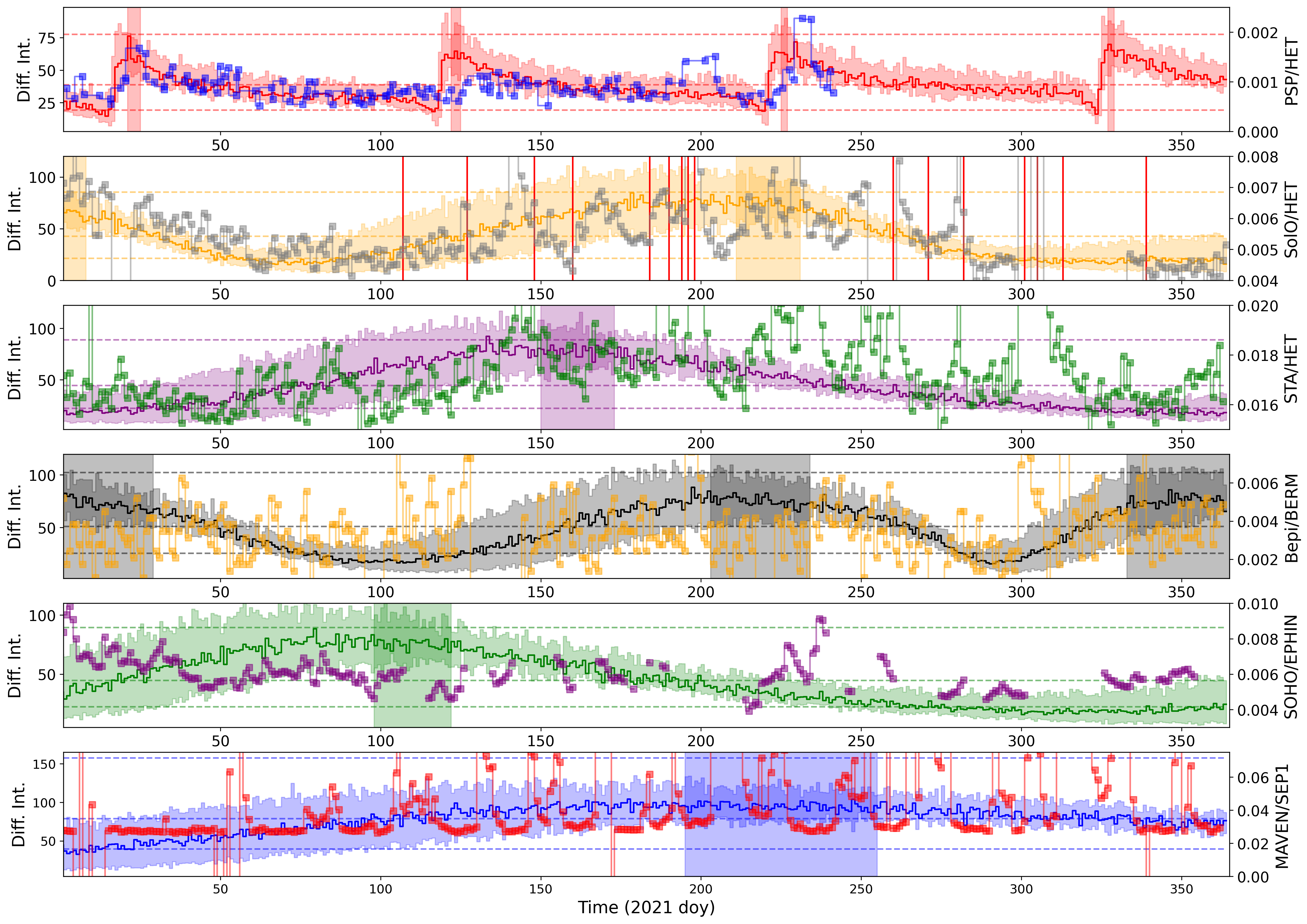}
    \caption{The lines and shaded uncertainty regions are modelling results for 4 MeV Jovian electrons along the selected spacecraft trajectories. These modelled intensities are in units of /m$^{2}$/s/sr/MeV. Spacecraft measurements are included as the scatter points with varying units: {PSP/HET: counts/s, SolO/HET: /cm$^2$/s/sr/MeV, STA/HET: /cm$^2$/s/sr/MeV, Bepi/BERM: counts/s, SOHO/EPHIN: /cm$^2$/s/sr/MeV, MAVEN/SEP1: /cm$^2$/s/sr/keV.} {The red vertical lines on the SolO panel indicate the occurrence of multi-spacecraft SEP events from the SERPENTINE SEP catalog.} }
    \label{fig:2021_simulations}
\end{figure*}

The top panel of Fig. \ref{fig:2D_trajectories} shows the trajectories, in terms of radial distance and longitude, of the selected spacecraft in 2021 as taken from \url{https://omniweb.gsfc.nasa.gov/coho/helios/heli.html}. The shaded regions indicate times when the different spacecraft are magnetically well-connected to the Jovian source{; see also Fig. \ref{fig:2021_simulations}}. {The color of the shading corresponds to that of the figure legend.} In the bottom panel we again show the trajectories (with a zoomed in image on the right), overlaid onto the simulated 4 MeV Jovian electron intensity as taken from \citet{Straussetal2011}. The motion of the Jovian source is also indicated on the figures and also included into the model calculations.\\

We now read the selected spacecraft trajectories, using a daily cadence, into the model of \citet{Straussetal2011} and compute the intensity of 4 MeV electrons. For the time being, we neglect any temporal changes in both the heliosphere (e.g. differing solar wind speeds) and in the transport coefficients. These results are shown in the different panels of Fig. \ref{fig:2021_simulations} for each spacecraft trajectory separately. The results using the nominal value of $\lambda = 0.1$ au are shown as the thick lines, while the shaded ranges indicate the uncertainty introduced by varying the transport coefficients by 25\%. Again, the vertical shaded regions indicate times when the spacecraft are magnetically well connected to Jupiter. \\

The measurement from the different spacecraft, summarized in the previous section, are included as the scatter points. As some of these measurements are not converted to a differential intensity (e.g. the counting rate from PSP) and as a thorough inter-calibration has not yet been performed (the energy bins of the spacecraft are also different), these intensity values are included on a separate axis on the right and scaled by eye to compare with the modelling results. While a more detailed quantitative comparison definitely needs to be done in the future, we are still able to qualitatively identify the general trends in both the simulation results and the measurements using the simplified approach outlined here. {For most of these detectors typical uncertainties are not routinely calculated and/or given. Statistical errors are usually quoted for some instruments but these are, at best, a lower limit for the actual uncertainty. As such we did not consider any data uncertainties in this work. } \\

The modelled profile at the position of SOHO (green curves) and STA (purple curves) show the expected $\sim 13$ month periodicity. The measurements at STA (green symbols) also reflect this profile, albeit with some scatter in the second half of the year due to multiple SEP events. SOHO (purple symbols), however, does not see the expected trend. It has to be noted though, that the electron flux observed by SOHO is dominated by SEPs except for the first third of the year. {The red vertical lines in the SolO panel indicate the occurrence of 25--40 MeV proton SEP events that were observed by at least two spacecraft, taken from the SERPENTINE SEP catalog \footnote{\url{ https://data.serpentine-h2020.eu/catalogs/sep-sc25/}}.} The absence of a clear variation in the first months could have been caused by unfavorable solar wind speeds and/or interaction regions preventing a magnetic connection to Jupiter. Due to Mars' slow orbital motion, the MAVEN simulations (blue curves) show a very weak Jovian electron signal. The corresponding measurements (red symbols) do not show any clear trend and are mostly consistent with constant fluxes. The faster orbiting spacecraft show quasi-periodically varying intensities. The SolO simulation (yellow curves) shows a periodicity of $\sim 250$ days, BepiColombo (black curves) $\sim 200$ days, and PSP (red curves) a period of $\sim 100$ days. The BepiColombo measurements (yellow symbols) show a hint of a Jovian signal, which may be enhanced by means of a more careful removal of SEP events. A very clear signal is seen in both SolO (gray symbols) and PSP (blue symbols) measurements. The simulated (and measured) PSP time profile is particularly interesting. Due to PSP's highly eccentric orbit, the time profile does not show sinusoidal variations, but rather a pulse-like profile. This can be understood by referring to the bottom right panel of Fig. \ref{fig:2D_trajectories}: The Jovian intensity profile at a spacecraft, in the ecliptic plane, is determined by both its radial and azimuthal position. While the radial changes are usually negligible for spacecraft at e.g. Earth's orbit, PSP's measurements are highly sensitive to these changes. In 2021, just before perihelion is reached, PSP measures a minimum Jovian intensity. This is because it is not magnetically connected to Jupiter, but it is also very near to the Sun and hence far from Jupiter. Just past perihelion, PSP is magnetically well-connected to Jupiter and measures a peak in the time profile, even though it is still near the Sun. The maximum Jovian intensity along PSP's orbit will only be measured during a magnetically well-connected aphelion. Such times are considered in the next section. \\

\section{Long term simulations}

\begin{figure*}
    \centering
\includegraphics[width=0.99\textwidth]{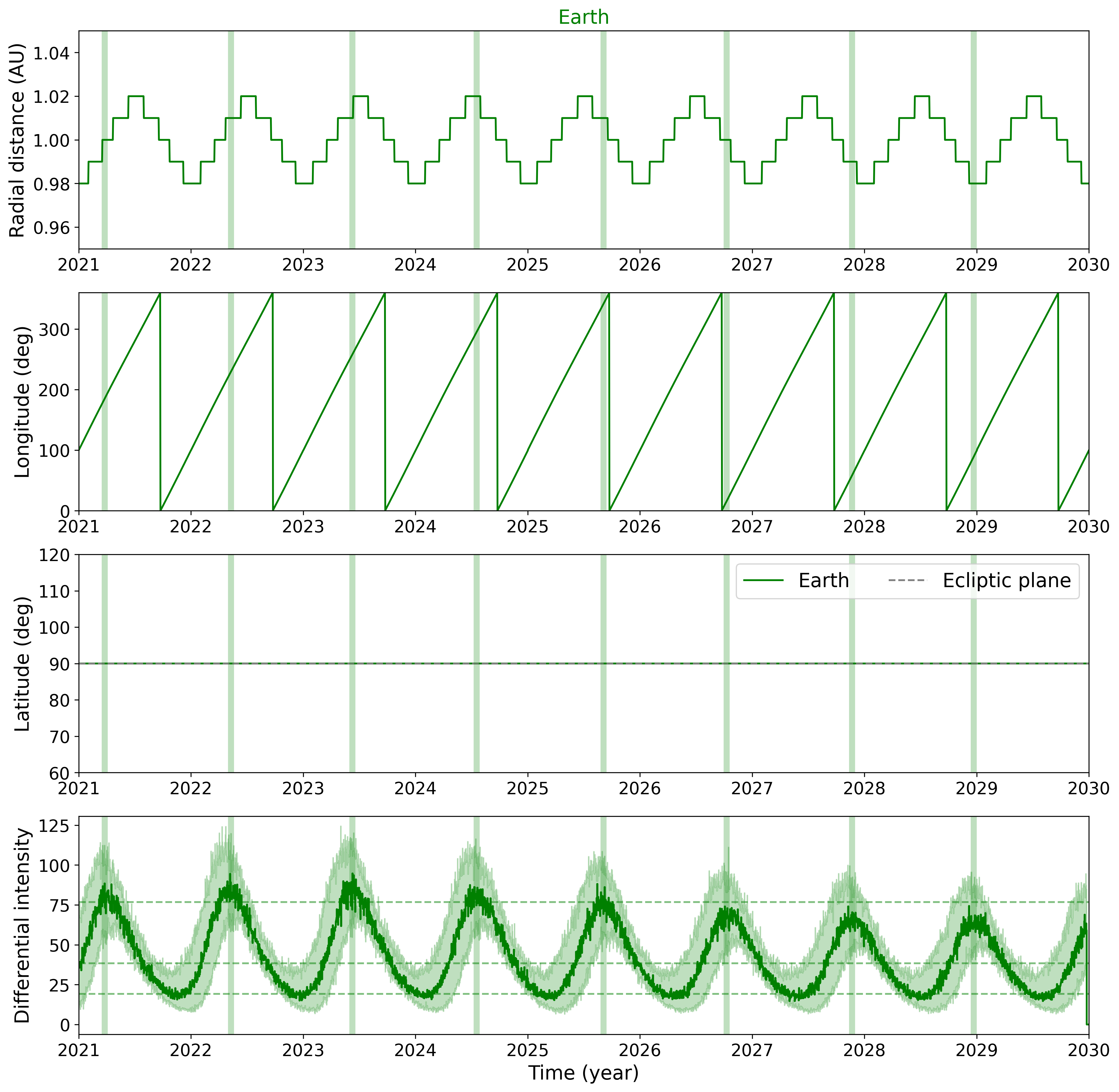}
    \caption{Long-term simulations of 4 MeV Jovian electrons at the position of Earth. Times of best magnetic connection between the spacecraft and Jupiter are also indicated by vertical shading.}
    \label{fig:long_term_earth}
\end{figure*}

\begin{figure*}
    \centering
\includegraphics[width=0.49\textwidth]{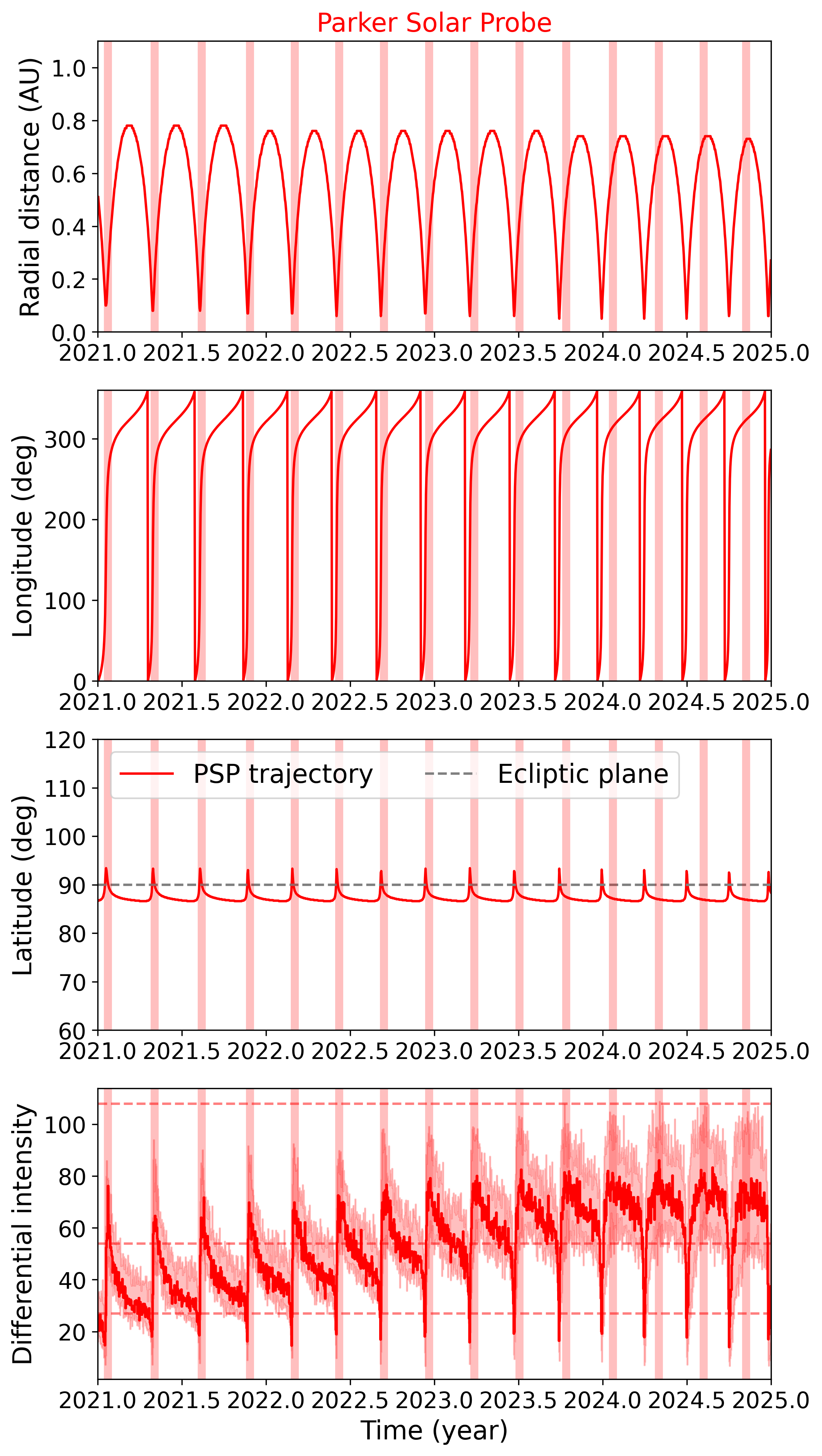}
\includegraphics[width=0.49\textwidth]{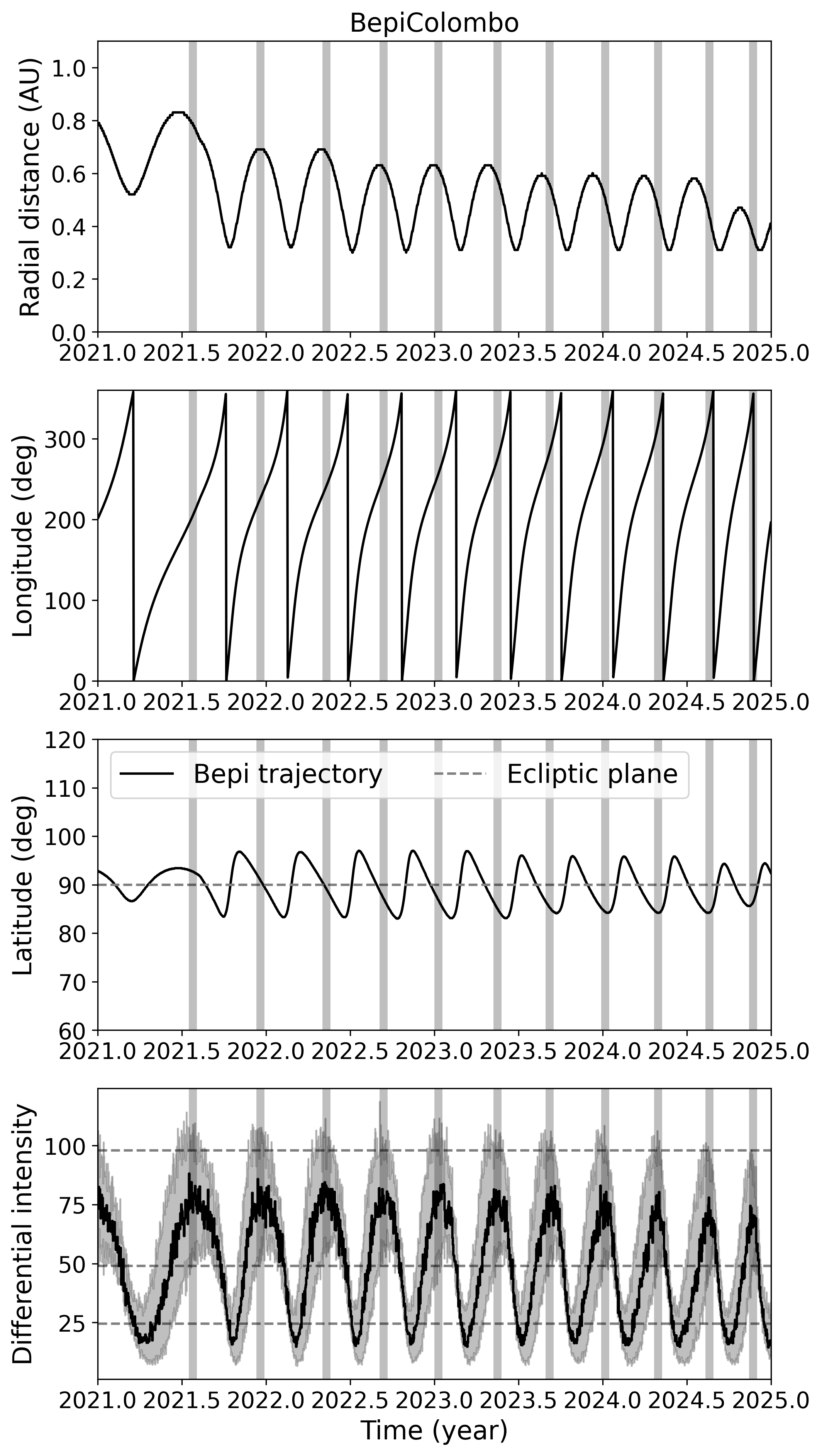}
    \caption{Long-term simulations of 4 MeV Jovian electrons at the position of PSP (left, in red) and BepiColombo (right, in black). Times of best magnetic connection between the spacecraft and Jupiter are also indicated by vertical shading.}
    \label{fig:long_term_psp_bepi}
\end{figure*}

\begin{figure*}
    \centering
\includegraphics[width=0.99\textwidth]{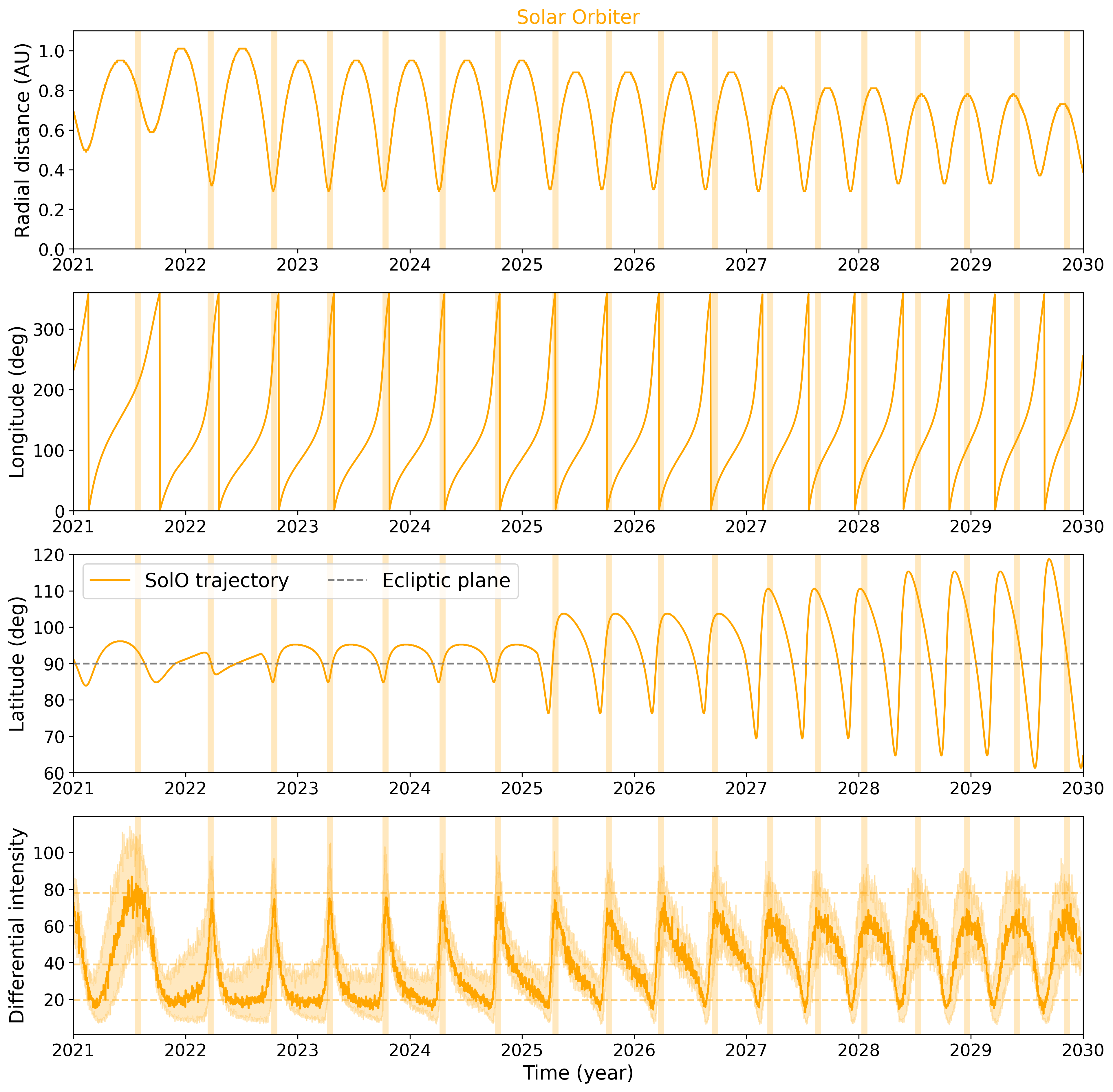}
    \caption{Long-term simulations of 4 MeV Jovian electrons at the position of SolO. Times of best magnetic connection between the spacecraft and Jupiter are also indicated by vertical shading.}
    \label{fig:long_term_solo}
\end{figure*}

Here, we repeat the calculations shown in the previous section, but use the position of selected spacecraft from 2021 to 2025 (for PSP and BepiColombo) and, for Earth and SolO, from 2021 to 2030. These trajectories and the resulting model calculations are shown in Figs. \ref{fig:long_term_earth}, \ref{fig:long_term_psp_bepi}, and \ref{fig:long_term_solo}.\\

The simulated intensity at Earth shows, again, the expected $\sim 13$ month periodicity, but interestingly also shows the effect of a changing radial position: The absolute maximum Jovian intensity for this time period, is predicted in 2023 when Earth is at aphelion during the best magnetic connection with Jupiter, while the maximum during late 2028, when Earth is at perihelion during best magnetic connection, is ~25\% lower. This is, of course, much smaller than the factor $\sim 3$ change that results from the changing magnetic connection (i.e. the difference from minimum to maximum during a $\sim 13$ month period).\\

The simulated profiles along PSP's future trajectory show the effect of a highly eccentric trajectory, with the Jovian intensity profile changing significantly from 2021 to e.g. 2024, where the best magnetic connection will be measured at aphelion. Similar changes are also predicted along the BepiColombo profile, albeit not as accentuated as for PSP.\\

For the simulations along SolO's trajectory, as shown in Fig. \ref{fig:long_term_solo}, we also include the effect of SolO's inclined orbit later in the mission. Up to $\sim$2028, expected Jovian intensity profiles display sharply peaked profiles similar to what is seen for PSP (Fig.~\ref{fig:long_term_psp_bepi}), for similar reasons. Beyond these times, the model yields more gradual, less pulse-like profiles more reminiscent of what is expected for, say, BepiColombo, with a periodicity of $\sim 135$ days. Care should, however, be taken in the interpretation of these latter profiles. During the period of interest, the solar activity cycle would be descending towards solar minimum, and as such the assumption made here of a latitudinally-constant solar wind speed would be less valid \cite[see, e.g.,][]{MccomasEA00}. For instance, higher solar wind speeds at higher latitudes would influence the heliospheric magnetic field winding angle, thereby altering the level (and period) of magnetic connectivity to the source. This latter phenomenon would be further complicated due to the possible presence of a Fisk-type heliospheric magnetic field \cite[see][and references therein]{SteynBurger20}, which could significantly affect the transport of Jovian electrons \cite[e.g.][]{SternalEA11}. Lastly, it should be noted that particle transport coefficients, due to both the solar cycle dependence of turbulence parameters and the differing nature of turbulence in the fast solar wind versus the slow solar wind \cite[e.g.][]{ForsythEA96,BavassanoEA00a,BavassanoEA00b}, would be expected to be considerably different to those applicable to slow solar wind conditions \cite[e.g.][]{EngelbrechtBurger13}. \\

\section{Historical, Upcoming, and Anticipated measurements}

\begin{figure*}
    \centering
\includegraphics[width=0.99\textwidth]{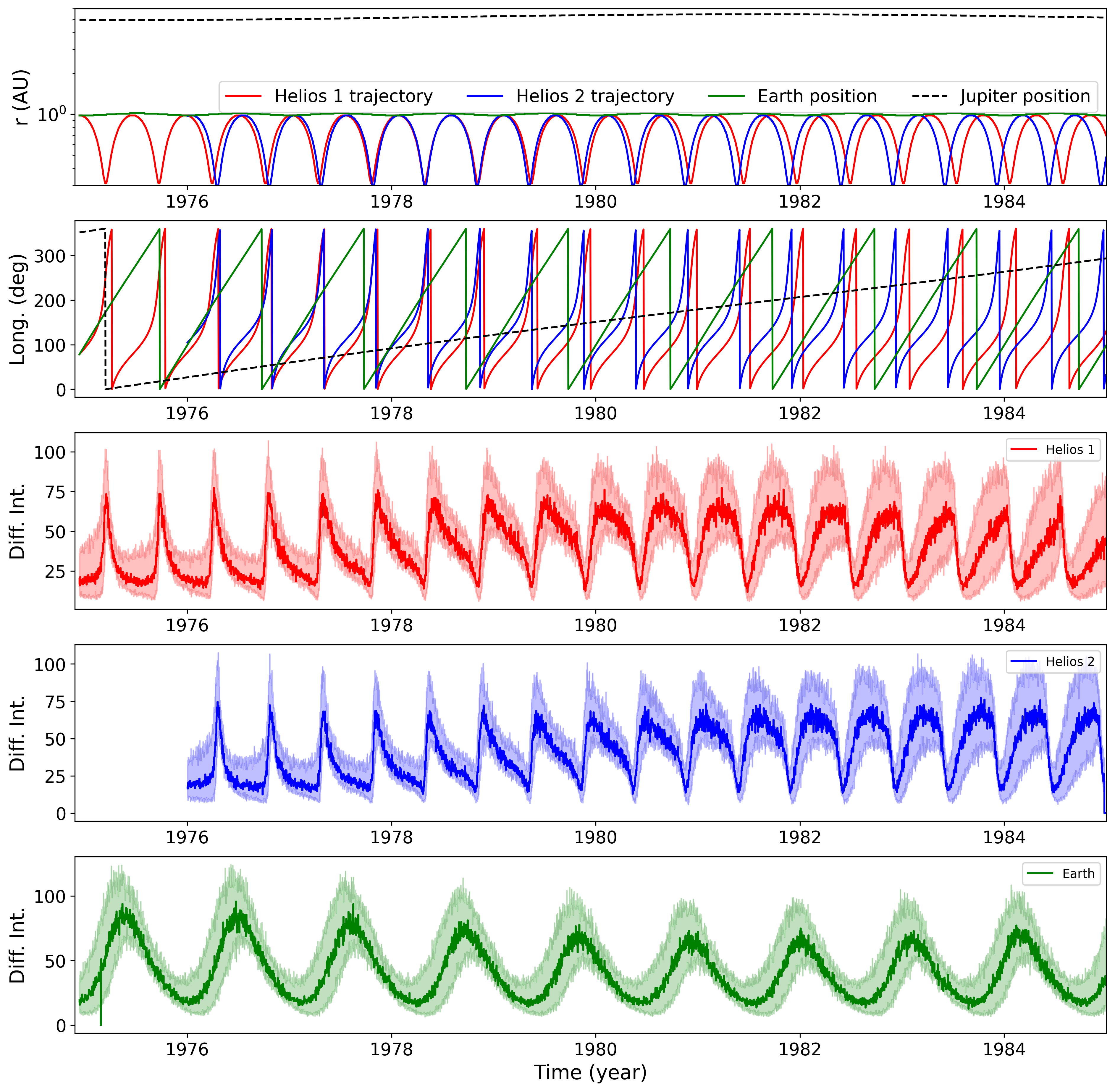}
    \caption{Historical simulations of 4 MeV Jovian electrons at the position of Earth (green), Helios 1 (red), and Helios 2 (blue) from 1975 until 1985 }
    \label{fig:legacy}
\end{figure*}

\subsection{IMP, Helios 1, and Helios 2}
As a last model application we solve the model to obtain the 4 MeV Jovian electron intensity along the trajectories of Earth (the position of the Interplanetary Monitoring Platform, IMP, spacecraft), Helios 1, and Helios 2, and present the results of these simulations in Fig. \ref{fig:legacy}.\\

\subsection{HEPD-02}

The second High Energy Particle Detector (HEPD-02), planned to be aboard the second China Seismo-Electromagnetic Satellite (CSES-02), is specifically designed to detect cosmic ray protons with energies ranging from 30-200 MeV and electrons with energies from 3-100 MeV, improving on the capabilities of HEP-01 aboard CSES-01 \cite[see, e.g.,][]{BartocciEA20,ScottiEA23}. Furthermore, information as to detected particle's pitch angles, albeit relative to the Earth's magnetic field, will also be available \citep{DeSantisEA21}. This mission is expected to launch in 2023.

\subsection{IMAP}

The upcoming Interstellar Mapping and Acceleration Probe (IMAP) mission \citep[][]{Mccomasetal2018} will continue relativistic electron measurements in the near-Earth solar wind with its High-energy Ion Telescope (HIT). Measurements from IMAP can serve as a future baseline value at Earth, and provide the linkage between ageing detectors on SOHO and ACE. IMAP is scheduled to be launched in 2025 into an orbit around the Sun–Earth L1 Lagrange point.

\subsection{ARTEMIS and Space Weather Needs}

It has been shown that MeV electron measurements can be exploited for space weather purposes \citep{Posner2007}. The premise of the  Relativistic Electron Alert System for Exploration (REleASE) algorithm is that all major SEP events release relativistic electrons at the onset, which arrive at 1 au tens of minutes before the more hazardous ions. This property of relativistic electrons is particularly relevant for the upcoming ARTEMIS era, during which the human presence on the Moon and in transit to and from Mars requires advance-warning of radiation hazards. The current implementation of REleASE,  HESPERIA (High Energy Solar Particle Events forecasting and Analysis) -- REleASE \citep[][]{MalandrakiCrosby2018} uses ACE/EPAM observations from the Earth-Sun Lagrange point 1 (L1) that are available in real time. The system is part of the SEP Scoreboard effort and as such is undergoing near-real-time testing. Current studies show that additional needs for relativistic electron observations from other vantage points exist. For example, human exploration of Mars would require a Mars upstream monitor  \cite[e.g.][]{PosnerStrauss2020,LeeEtal2022}. There are considerations to equip the transfer vehicle humans would use on their outward journey to Mars and return to Earth with relativistic electron detectors. Other space weather missions under consideration would lead to the Sun–Earth Lagrange points L4 \citep[][]{Posneretal2021} and L5 \citep[][]{Akiokaetal2005} and could be equipped with MeV electron instrumentation, although their use is not directly driven by space weather needs. Reliable, high-quality MeV electron measurements at multiple, well thought-out locations in the heliosphere would not only benefit human exploration, but would support in-depth Jovian electron studies. From a physics point-of-view, Jovian electrons and SEP electrons propagate in exactly the same fashion through the turbulent interplanetary medium. One can therefore make the argument that dedicated space weather missions, and their users, will also benefit from better understanding electron transport conditions in the heliosphere, and their potential variability, that we reach through studying Jovian electron transport.\\

\subsection{{Jupiter Icy Moons Explorer}}

{JUpiter ICy moons Explorer (Juice) is an ESA-mission, launched on 14 April 2023, to study the Jovian system. On board is the PEP (Particle Environment Package), a particle spectrometer to measure neutral and charged particles in the Jupiter system, including low energy Jovian electrons and other energetic particle populations \citep[][]{Pintoetal2020,Barabashetal2022}. Measurements taken during this mission could be used to revisit the form of the Jovian electron source spectrum employed in transport studies, providing an invaluable additional constraint for future modelling endeavours.}

\section{Discussion}

In this paper we discuss how multi-spacecraft Jovian electron measurements can be used to study the fundamental transport processes that ultimately determine their intensities throughout the heliosphere. We show an example of such a study using a simplified model that assumes relatively {\it ad hoc} transport coefficients in a time-stationary heliosphere. Such a study would need multiple sets of observations at various locations, taken during overlapping periods of time so as to disentangle solar-cycle and other temporally-dependent influences on transport coefficients \cite[see, e.g.,][]{Engelbrecht_2022}. Careful comparisons of the results of such studies with theory would not only improve our understanding of the theoretical underpinnings of electron diffusion, but could also give indirect measures of hard-to-measure turbulence quantities pertaining to the dissipation range of solar wind turbulence, as for example it is known that low-energy electron parallel mean free paths are strongly dependent on both the dissipation range onset wavenumber and spectral index \cite[see, e.g.,][]{EngelbrechtBurger13}. This would also have the potential to provide a novel, albeit indirect measure of the solar-cycle dependence of these quantities.\\

We present a comparison between different spacecraft measurements of Jovian electrons and simulation results specifically for 2021. The model is able to capture the observed Jovian electron features very well, especially the somewhat peculiar time profiles that were observed by PSP. The PSP results confirm the ability of Jovian electrons to reach small heliocentric distances without being completely impeded by the outward moving solar wind. However, for the majority of the time periods during which PSP was nominally magnetically connected to Jupiter, including the second perihelion of 2021, IS$\odot$IS did not observe a coincident Jovian electron enhancement as predicted by the model. Several potential reasons for the absence of Jovian electron enhancements during these periods are explored in \cite{mitchell2022}, including impediments from stream interaction regions (SIRs), modulation of the Jovian source by the solar wind dynamic pressure, and the obstruction of the flow of Jovian electrons by the heliospheric current sheet (HCS) when the spacecraft and Jupiter lie on opposite sides of the HCS. The later time profile along PSP's trajectory is predicted to differ substantially from that measured in 2021 due to PSP's highly eccentric orbit. The maximum intensities measured during each orbit are predicted to occur near aphelion later in the mission, resulting in a more constant profile with a deep depression near perihelion. The predicted intensity profile from 2022 to $\sim 2025$ at SolO's position is particularly interesting to study due to a sharply peaked profile near perihelion.\\

We note that some spacecraft measurements, e.g. those of SOHO and BepiColombo, do not show a clear Jovian signal. {It is not yet entirely clear as to why this is the case, and future studies, combining both numerical transport modelling and further data analyses, will investigate this phenomenon in more depth.} {We also note that for BepiColombo, the area of the entrance to the top detector is very small, 0.5 mm$^2$ \citep{Pinto2022} and the field of view is $40^{\circ}$, which may be contributing to a reduce rate of detection of Jovian particles, and which statistic uncertainty will be considered in future analysis.}\\

While the proposed multi-spacecraft Jovian electron studies are compelling, long-term historical data-sets can also yield novel insights into phenomena observed in the newer observations. Here we show, as an example, the expected profiles along selected spacecraft trajectories from 1975 -- 1985. These measurements show the same features as their modern-day counterparts. Similar multi-spacecraft Jovian electron studies would therefore also be possible by using historical data, and would accordingly require more careful modelling of the temporal dependence of these particle's transport coefficients and the turbulence quantities they depend upon. Although this has been done to a degree in various galactic cosmic ray proton studies, to date no such physics-first studies pertaining to the time-dependent transport of MeV electrons exist.\\

Several planned or proposed future missions will have the capability to measure Jovian electrons. These valuable measurements can help to constrain the transport processes of MeV electrons through the turbulent interplanetary medium. We argue that such studies will also be beneficial to space weather studies, e.g. forecasting SEP events, as well as other human space-borne endeavours, and that dedicated space weather missions could, and should, also be leveraged to measure Jovian electrons.\\

The modelling endeavours discussed in this paper can be refined and improved in several ways. Firstly, the models and parameters employed for large-scale plasma quantities, such as the solar wind speed and heliospheric magnetic field, can be brought closer to spacecraft observations in terms of, say, solar cycle dependencies, {to further investigate the relative significance of temporal effects}. Jovian transport coefficients can also, for instance, be modelled in a more fundamental way, incorporating the latest results from various particle scattering theories \cite[e.g.][]{Shalchi20,EngelbrechtEA22b}, so as to improve the ability of the model in terms of extrapolation as well. This has already been done for galactic electrons \cite[e.g.][]{DempersEngelbrecht20,Engelbrecht19}, and the results of such studies can be applied to the study of Jovians. Such an approach, however, would require careful modelling of turbulence quantities that such transport coefficients are functions of. For this, too, several successful turbulence transport models exist \cite[see, e.g., the reviews of][]{OughtonEngelbrecht20,AdhikariEA21,FraternaleEA22} that can readily be applied to such studies.\\

\acknowledgments

This work is based on the research supported in part by the National Research Foundation of South Africa (NRF grant numbers RA170929263913, SRUG220322419, and 137793). Opinions expressed and conclusions arrived at are those of the authors and are not necessarily to be attributed to the NRF. The responsibility of the contents of this work is with the authors. N.D. is grateful for support by the Academy of Finland (SHOCKSEE, grant No.\ 346902). B.S.-C. acknowledges support through UK-STFC Ernest Rutherford Fellowship ST/V004115/1 and STFC grant ST/Y000439/1. The EPD and this work was supported by the German Space Agency (Deutsches Zentrum f\"ur Luft- und Raumfahrt, e.V., DLR) under grant number 50OT2002 to the Kiel University. Figures prepared with Matplotlib \citep{Hunter:2007} and certain calculations done with NumPy \citep{Harrisetal2020}.


\bibliography{ref}

\appendix
\section{Filter method of EPHIN data}\label{sec:AppEphinFilterMethod}

\begin{figure}
    \centering
    \includegraphics[width=0.8\textwidth]{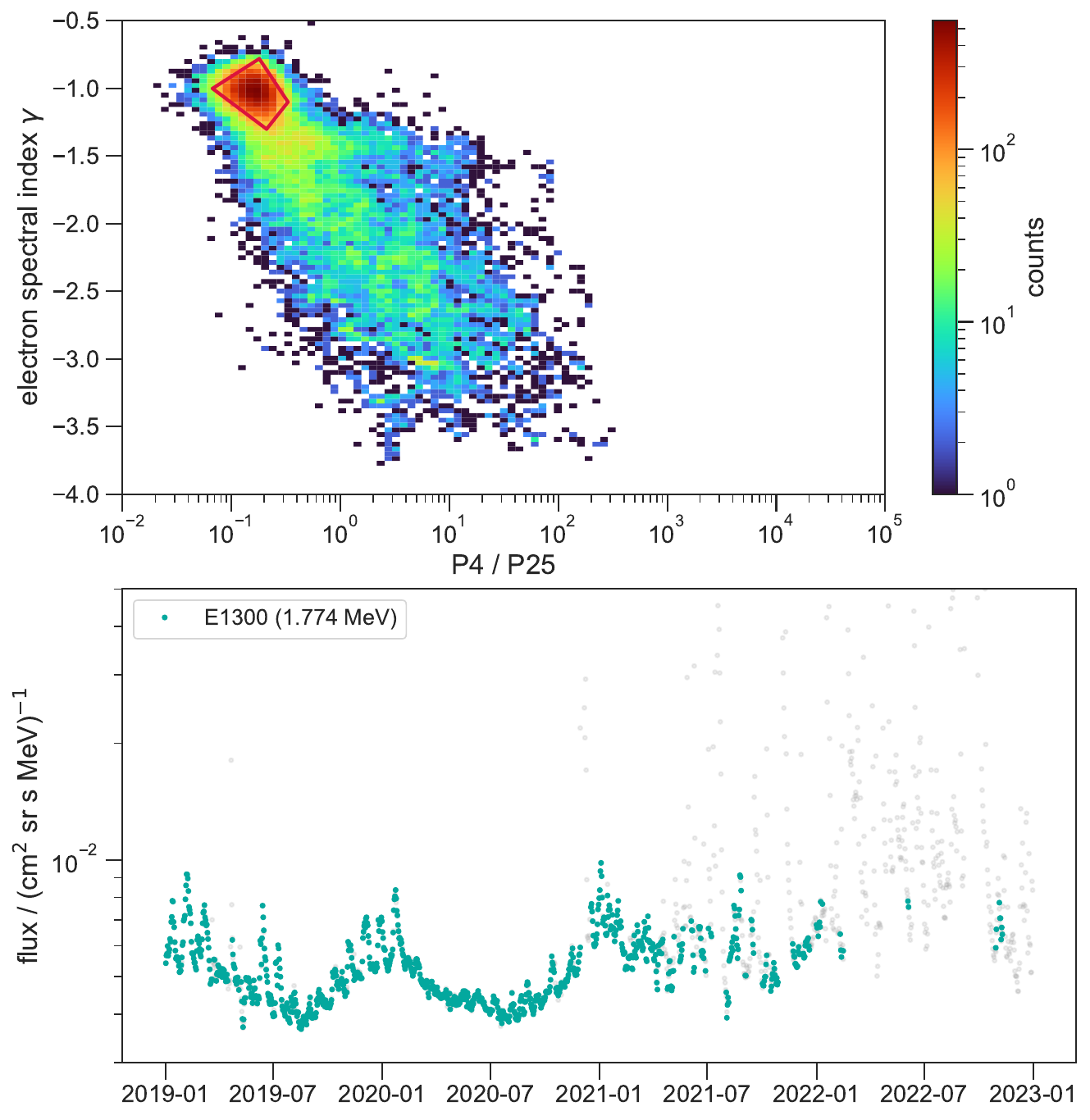}
    \caption{{Electron spectral index plotted against the ratio of proton channels P4 and P25 (top) and resulting time profile of E1300 flux from 2019 to 2023 (bottom).}}
    \label{fig:ephinSepFilter}
\end{figure}

{To better see the modulation of the electrons in the time profiles, it is necessary to filter out the SEP events. In order to do this we used two electron channels, the E150 ($E_{eff} = 0.5$~MeV) and E1300 ($E_{eff} = 1.8$~MeV) channels, and determined the spectral slope for an assumed power-law spectrum on hourly averages. In the top panel of fig.~\ref{fig:ephinSepFilter} we have plotted these spectral indices against the ratios of the proton P4 and P25 count rates. The P4~/~P25 ratio serves as a proxy of the spectral index of the protons. Since the energy spectrum is generally softer during SEP events, we assume that electrons outside of the red box shown in the figure are of solar origin. The data used in this study consist only of the electrons inside the selection box, reducing the contribution of SEP events.  The bottom panel of fig.~\ref{fig:ephinSepFilter} shows the time profile of the E1300 electron flux from 2019 to 2023 as a daily average. The teal and grey dots show the filtered (i.e. red selection box) and unfiltered data, respectively. }

\end{document}